\documentclass[runningheads]{llncs}
\usepackage[T1]{fontenc}
\usepackage{graphicx}
\usepackage{amsmath,amsfonts}
\usepackage{xcolor}
\usepackage{textcomp}
\usepackage{url}
\usepackage[hidelinks]{hyperref}
\usepackage{listings}
\usepackage{cleveref}
\begin{document}
\title{Guess My Weight: Profiled Side-Channel Recovery of Floating-Point Neural-Network Weights}
\titlerunning{Profiled Side-Channel Recovery of Floating-Point Neural-Network Weights}
%
\author{Timon Lumír Fillo\inst{1}\and
Ján Mikulec\inst{1}\and
Anubhab Baksi\inst{2}\and
Jakub Breier\inst{3}\and
Xiaolu Hou\inst{1}
}
\institute{Slovak University of Technology, Bratislava, Slovakia\\
\email{xfillo@stuba.sk, jan.mikulec@stuba.sk, houxiaolu.email@gmail.com}\and
Indian Institute of Technology Bhilai, India\\\email{anubhab.baksi@eit.lth.se}\and
TTControl GmbH, Vienna, Austria\\\email{jbreier@jbreier.com}
}
\authorrunning{T. L. Fillo et al.}
%
\maketitle              
\begin{abstract}

Neural-network parameters deployed on embedded devices may be exposed through physical side-channel leakage during inference. 
Existing side-channel attacks on floating-point neural-network parameters have often targeted reduced numerical precision, while recovering the complete IEEE-754 representation remains considerably more challenging because of the large and structured 32-bit candidate space. 
We present a profiled template attack for bit-exact recovery of an IEEE-754 single-precision neural-network weight from power measurements. The attack targets the floating-point multiplication between a known input and a first-layer weight. During profiling, multivariate Gaussian templates are learned from randomized network configurations using Hamming-weight classes of the multiplication result, while the remaining network parameters act as nuisance variables. To efficiently search the structured 32-bit floating-point candidate space, we use a hierarchical coarse-to-fine-to-exact procedure that progressively increases both the numerical and leakage-model resolution. Experiments on a ChipWhisperer-Lite with an Arm Cortex-M4 demonstrate recovery of the exact \texttt{float32} representation of the target weight. In the evaluated setting, the attack reaches a bit-exact success rate of 99\% with 171 traces and 100\% from 263 traces onward. These results demonstrate that profiling can enable practical full-precision extraction of floating-point neural-network parameters from physical leakage.

\end{abstract}
\section{Introduction}

Neural-network models deployed on embedded and edge devices often represent valuable intellectual property, as their trained parameters encode knowledge obtained through costly data collection and optimization.
At the same time, executing neural-network inference on physical devices can expose model information through side-channel leakage~\cite{chabanne2021parasite,hashemi2022hwgn}.
Power consumption and electromagnetic emanations generated during arithmetic operations may depend on the processed data and model parameters, enabling an adversary with physical measurement capabilities to infer information about the internal model.

Prior work has demonstrated that side-channel leakage can reveal neural-network architectures, processed inputs, and model parameters~\cite{batina2019csi,wei2018know,joud2022practical,nozaki2024power}.
Parameter recovery attacks have predominantly relied on non-profiled techniques such as correlation power or electromagnetic analysis, while previous applications of template attacks to neural-network implementations have mainly considered input recovery.
Several existing side-channel attacks on neural-network parameters assume a limited numerical precision when recovering floating-point weights~\cite{batina2019csi,puvskavc2025make}.
More recent approaches have addressed higher-precision floating-point parameter recovery using iterative or IEEE-754-aware correlation-based techniques~\cite{joud2022practical,nozaki2024power}.
Bit-exact recovery has been shown in~\cite{breier2021sniff} where the authors utilized a fault injection attack to reverse-engineer the last layer of the network.
In contrast, this work investigates exact 32-bit weight recovery in a profiled setting, where candidate hypotheses are evaluated using learned multivariate leakage templates rather than direct correlation with the measured traces.
To the best of our knowledge, a profiled template attack specifically targeting bit-exact 32-bit floating-point neural-network weight recovery has not previously been investigated.

We investigate a profiled template attack that targets the multiplication between a known network input and the first weight of the first layer.
During profiling, traces are collected across randomized network configurations for which the corresponding profiling value of the attacked weight is known, while the remaining parameters are treated as nuisance variables.
The resulting templates characterize the leakage of the floating-point intermediate $\mathrm{float32}(wx)$ and are subsequently used to evaluate candidate weights on a fixed target network.
Weight hypotheses are ranked by their accumulated template likelihood over multiple attack traces, allowing the unknown parameter to be recovered without directly correlating the measured traces with hypothetical leakage values.

A key challenge is that the target parameter is represented as a 32-bit IEEE-754 floating-point value, resulting in a large and finely structured candidate space.
To make exact recovery practical, we use a hierarchical coarse-to-fine-to-exact search that progressively increases the resolution of the leakage model.
The initial stages use only the most significant parts of the floating-point representation to identify promising regions, while the final stage evaluates neighboring representable \texttt{float32} values using the complete 32-bit leakage model.
This strategy combines efficient candidate reduction with bit-exact recovery of the target weight.

The main contributions of this work are as follows.
\begin{itemize}
\vspace{-0.2cm}
    \item We present a profiled template attack for recovering an IEEE-754 single-precision neural-network weight from physical side-channel measurements.
\item We construct leakage templates across randomized network configurations that can be applied to recover an unknown target weight with no knowledge of the remaining model parameters.
    \item We introduce a hierarchical coarse-to-fine-to-exact search that enables practical and bit-exact 32-bit weight recovery.
    \item We experimentally demonstrate bit-exact recovery of the targeted \texttt{float32} weight and quantify the attack efficiency through a success-rate analysis, achieving 99\% success with 171 attack traces and 100\% success from 263 traces onward in the evaluated setting.
\end{itemize}


\section{Background and Related Work}
\label{sec:background}

\subsection{Side-Channel Analysis Attacks on Neural Networks}
\label{sec:scaonnn}
During neural-network~\cite{goodfellow2016deep} inference, each layer transforms its input activations through operations such as weighted multiplications, accumulations, and nonlinear activation functions.
In a fully connected layer, for example, the output of a neuron is obtained from a weighted sum of its inputs followed by an activation function.
Since these computations repeatedly process network parameters and intermediate activations, their execution may produce data-dependent physical leakage.

Side-channel analysis (SCA) can exploit such leakage to infer sensitive information about a neural-network implementation.
Depending on the attack objective, the targeted information may include the network architecture, trained parameters, or data processed by the model.
An adversary can observe physical characteristics of the implementation, including execution time, power consumption, or electromagnetic (EM) emissions, while the network performs inference or training.
Previous studies have shown that timing leakage can reveal information about activation functions~\cite{batina2019csi}, while power and EM measurements can disclose information about network inputs~\cite{batina2019poster} or structural properties of the model, such as the types of executed layers~\cite{yan2023mercury}.

Particularly relevant to parameter extraction are attacks based on correlation power analysis (CPA)~\cite{batina2019csi,puvskavc2025make,lehocky2025side}.
In CPA, an attacker generates hypothetical intermediate values for different candidate parameters and maps them to predicted leakage values using a suitable leakage model.
The predicted leakage is then statistically correlated with the measured power or EM traces.
A candidate producing a consistently high correlation is considered more likely to correspond to the secret parameter.
The effectiveness of this approach relies on the relationship between the data manipulated during neural-network computation and the resulting physical leakage.

\subsection{Template Attacks}
\label{sec:template-attacks}

Template attacks are a class of profiled side-channel attacks in which an adversary first characterizes the leakage behavior of a device under controlled conditions and subsequently uses the resulting statistical model to recover an unknown sensitive value~\cite{chari2002template}.
In the classical setting, profiling traces with known intermediate values are used to construct multivariate leakage distributions at selected points of interest.
During the attack phase, candidate hypotheses are evaluated according to their likelihood under these templates.

For each leakage class $c$, the leakage vector $\mathbf{t}$ is commonly modeled using a multivariate Gaussian distribution,
\begin{equation*}
    \mathbf{t} \mid c
    \sim
    \mathcal{N}
    \left(
        \boldsymbol{\mu}_c,
        \boldsymbol{\Sigma}_c
    \right),
\end{equation*}
where $\boldsymbol{\mu}_c$ and $\boldsymbol{\Sigma}_c$ denote the class-dependent mean vector and covariance matrix, respectively.

Template attacks have also been considered in the context of neural-network implementations.
Wei et al.~\cite{wei2018know} applied a power template attack to an FPGA-based convolutional neural-network accelerator with the objective of reconstructing the network input from side-channel measurements.
Their attack targets information related to the processed image rather than the model parameters.

To the best of our knowledge, previous applications of classical template attacks to neural-network implementations have primarily focused on input recovery, and we are not aware of prior work applying a profiled template attack specifically to the recovery of floating-point neural-network weights.
This motivates the study of template-based parameter recovery considered in this work.

\subsection{IEEE-754 Floating-Point Representation}
\label{sec:ieee754}

The target implementation represents neural-network weights, inputs, and intermediate values using IEEE-754 single-precision floating-point numbers~\cite{IEEE754}.
A 32-bit floating-point value consists of one sign bit, eight exponent bits, and twenty-three fraction bits,
\begin{equation*}
    b_{31}\ldots b_0 =
    \underbrace{b_{31}}_{\text{sign}}
    \underbrace{b_{30}\ldots b_{23}}_{\text{exponent}}
    \underbrace{b_{22}\ldots b_0}_{\text{fraction}}.
\end{equation*}
The IEEE-754 standard refers to the lower $23$ stored bits as the fraction field.
These bits are also commonly referred to informally as the mantissa bits; in this work, we use the term \emph{fraction bits} throughout.

For a normalized value, the represented number is
\begin{equation*}
    x =
    (-1)^s
    \left(
        1 + \frac{M}{2^{23}}
    \right)
    2^{E-127},
\end{equation*}
where $s$ is the sign bit, $E$ denotes the unsigned exponent field, and $M$ is the integer represented by the fraction bits.

This representation is relevant to side-channel analysis because arithmetic operations manipulate data whose binary encoding depends on the sign, exponent, and fraction fields.
In particular, numerically close floating-point values may differ only in a small number of fraction bits, while larger numerical changes may also modify the exponent field.
These properties influence the leakage models used for recovering floating-point neural-network parameters.

The hierarchical recovery strategy in this work exploits this structure by increasing the number of considered fraction bits across three stages.
A reduced-bit representation is used for the initial search to provide robustness over a wide candidate range, while progressively more bits are included during refinement to distinguish increasingly close floating-point candidates.

\section{Threat Model and Experimental Setup}
\label{sec:experimental-setup}

\subsection{Threat Model}
\label{sec:threat-model}

We consider a profiled adversary.
The adversary knows the network architecture and the position of the targeted multiplication in the forward pass, can choose the network input $x$, and can measure the power consumption of the target device during inference.
For profiling, the adversary has a device of the same type that can be programmed with weights of the adversary's choice, so the targeted weight is known for every profiling trace.
On the attacked device, the weights are fixed and unknown.
In our experiments, the profiling and attack traces are acquired on the same physical board, which is re-programmed between the two phases.

\subsection{Target Neural Network}
\label{sec:target-network}

We consider a feed-forward neural network implemented in software on the target microcontroller.
The network consists of four layers with $7$, $5$, $4$, and $3$ neurons, respectively.
The first layer represents the input layer, while the following three layers are fully connected.
The hidden layers use the ReLU activation function, whereas the output layer applies a sigmoid activation.
All weights and intermediate values are represented as IEEE-754 single-precision floating-point numbers.

The objective of the attack is to recover the first weight of the first fully connected layer,
\begin{equation*}
    w^\star = W^{(1)}_{0,0}.
\end{equation*}
In the fixed target network used during the attack phase, this weight is set to $1.43$ in the firmware source.
At run time, the device stores the nearest \texttt{float32} value, $1.42999994754791\ldots$ (bit pattern \texttt{0x3FB70A3D}).

The targeted weight is particularly suitable for a first side-channel analysis because it is involved in a multiplication with an attacker-chosen input.
For each inference, the attacker supplies a floating-point value $x$, which is assigned to the first neuron of the input layer.
The remaining input neurons keep their fixed initial activation of $0.5$.
During the forward pass, the first neuron of the first hidden layer therefore computes a term of the form
\begin{equation*}
    z_0^{(1)}
    =
    b_0^{(1)}
    +
    w^\star x
    +
    \sum_{j=1}^{6} W^{(1)}_{0,j} a_j^{(0)},
\end{equation*}
where $a_j^{(0)}$ denotes the activation of input neuron $j$ and $b_0^{(1)}$ is the neuron bias.
In the implemented network, the biases are initialized to zero.
The forward pass processes the neurons in order, starting with the first input of the first neuron.
The first multiplication of the whole inference is consequently
\begin{equation*}
    v = \mathrm{float32}(w^\star x).
\end{equation*}
The target microcontroller (STM32F303) has a hardware floating-point unit, but the firmware is compiled with the ChipWhisperer default \texttt{-mfloat-abi=soft}.
Every floating-point multiplication is therefore executed by a software library routine that processes the sign, exponent, and fraction of the operands with integer instructions.

This intermediate value is the leakage target considered throughout this work.
Since $x$ is known to the attacker, each candidate value $\hat{w}$ induces a hypothetical intermediate
\begin{equation*}
    \hat{v}
    =
    \mathrm{float32}(\hat{w}x),
\end{equation*}
from which a corresponding leakage class can be predicted.
The attack therefore does not attempt to observe the weight value directly.
Instead, it exploits the data-dependent leakage generated while the processor evaluates the floating-point multiplication between the unknown weight and the known input.

Targeting the first weight also limits uncertainty in the processed operand.
For weights in later layers, the corresponding input activation depends on previous weighted sums and nonlinear activation functions and is therefore generally unknown unless earlier network parameters are already available.
In contrast, for $W^{(1)}_{0,0}$ one operand of the multiplication is directly selected by the attacker.
This makes the first weight a natural target for evaluating the feasibility of profiled template-based neural-network parameter recovery.

Although the analysis focuses on a single weight, the complete forward pass is executed during every measurement.
Consequently, the measured power traces contain leakage not only from the targeted multiplication but also from the remaining multiplications, additions, activation functions, and control operations of the network.
During profiling, the non-target weights are varied together with the target weight and are not explicitly incorporated into the leakage model.
They therefore act as nuisance parameters whose contribution must be tolerated by the profiling procedure.

\subsection{Profiling Scenario}
\label{sec:profiling-scenario}

The profiling phase is used to characterize the side-channel leakage of the targeted floating-point multiplication.
For this purpose, all network weights are drawn at random from the interval $[-2,2]$, and a new random configuration is generated every $100$ traces (Section~\ref{sec:measurement-platform}).

The targeted weight $W^{(1)}_{0,0}$ is known during profiling.
For every acquired trace, the corresponding input value $x$ and the current value of the targeted weight are stored together with the power measurement.
This information allows the intermediate product
\begin{equation*}
    v = \mathrm{float32}(W^{(1)}_{0,0} x)
\end{equation*}
to be reconstructed exactly for the profiling measurements.

The remaining network weights are also randomized, while the biases stay zero in both phases.
However, the non-target weights are not used by the leakage model and are not required during template construction.
They therefore act as nuisance parameters whose influence is implicitly captured in the measured power traces.
This profiling setting is intended to reduce dependence on one particular fixed network configuration.

The profiling dataset contains $50{,}000$ traces.
The network configuration is refreshed once every $100$ traces.
Consequently, the dataset contains $500$ distinct randomized network configurations.
Within each configuration, the weight values remain fixed while different input values are applied.

Let the profiling dataset be denoted by
\begin{equation*}
    \mathcal{D}_{\mathrm{prof}}
    =
    \left\{
        \left(
            x_i,
            w_i,
            \mathbf{T}_i
        \right)
    \right\}_{i=1}^{N_{\mathrm{prof}}},
\end{equation*}
where $x_i$ is the known network input, $w_i$ is the known value of the targeted weight, and $\mathbf{T}_i$ is the corresponding measured power trace.
In our experiments,
\begin{equation*}
    N_{\mathrm{prof}} = 50{,}000.
\end{equation*}

For each profiling measurement, the known pair $(x_i,w_i)$ is used to compute the hypothetical intermediate processed by the target device.
The corresponding leakage class is then derived from the IEEE-754 representation of the product.
These labeled profiling traces are subsequently used for point-of-interest selection and for construction of the multivariate Gaussian templates described in Section~\ref{sec:template-attack}.

The profiling phase is performed independently from the attack phase.
In particular, the fixed target-network weight used during the attack is not required for template construction.
The templates characterize the leakage associated with classes of floating-point multiplication results rather than with individual weight values.
As a result, the same profiling model can be evaluated against candidate weights of an independently configured target network.

\subsection{Attack Scenario}
\label{sec:attack-scenario}

During the attack phase, the neural-network configuration is fixed and the value of the targeted weight $W^{(1)}_{0,0}$ is unknown to the attacker.
The attacker can provide the input value $x$ and acquire the corresponding power trace generated during network inference.
Unlike the profiling phase, the value of the targeted weight is not included in the attack-trace metadata.

The attack dataset is represented as
\begin{equation*}
    \mathcal{D}_{\mathrm{attack}}
    =
    \left\{
        \left(
            x_i,
            \mathbf{T}_i
        \right)
    \right\}_{i=1}^{N_{\mathrm{attack}}},
\end{equation*}
where $x_i$ denotes the known input value and $\mathbf{T}_i$ is the corresponding measured power trace.
The unknown parameter $w^\star$ remains constant across all traces because the same target network is used throughout the attack experiment.

For a candidate weight $\hat{w}$, the attacker computes the hypothetical intermediate
\begin{equation*}
    \hat{v}_i
    =
    \mathrm{float32}(\hat{w}x_i).
\end{equation*}
The IEEE-754 representation of $\hat{v}_i$ is then mapped to a leakage class according to the leakage model established during profiling.
The measured trace $\mathbf{T}_i$ is compared with the template corresponding to this hypothetical class.

Each attack trace therefore provides statistical evidence about the correctness of the candidate weight.
Evidence from multiple traces is accumulated using the template log-likelihood,
\begin{equation*}
    L(\hat{w})
    =
    \sum_{i=1}^{N_{\mathrm{attack}}}
    \log
    p
    \left(
        \mathbf{t}_i
        \mid
        c_i(\hat{w})
    \right),
\end{equation*}
where $\mathbf{t}_i$ denotes the samples of trace $\mathbf{T}_i$ at the selected points of interest and $c_i(\hat{w})$ is the hypothetical leakage class predicted for candidate $\hat{w}$.

The candidate with the highest accumulated likelihood is considered the most plausible value of the targeted weight.
To avoid evaluating the complete IEEE-754 weight space directly, the attack applies the hierarchical coarse-to-fine-to-exact search described in Section~\ref{sec:hierarchical-recovery}.
The search progressively reduces the candidate space while increasing the precision of the leakage model.

For experimental validation, the true weight value is known to the evaluator from the fixed network configuration.
This value is used only to assess whether the attack successfully recovered the targeted parameter and is not used during candidate scoring or template matching.

\subsection{Measurement Platform}
\label{sec:measurement-platform}

Power measurements are acquired with a ChipWhisperer-Lite and its integrated STM32F303 (Arm Cortex-M4) target (\texttt{CWLITEARM}).
The firmware is programmed through the STM32 serial bootloader.
The scope uses the ChipWhisperer default settings: the target is clocked at $7.37$\,MHz, the ADC samples at four times the target clock ($29.48$\,MS/s), and the gain is $25$\,dB.
Communication between the host and the target is implemented using the ChipWhisperer SimpleSerial protocol.

The capture script uses a single fixed acquisition window for every inference.
Each trace contains $24{,}400$ samples (about $6{,}100$ target clock cycles), close to the capture-buffer limit of the ChipWhisperer-Lite.
No sample decimation is applied, i.e., the decimation factor is set to one.
Before collecting either the profiling or attack dataset, $50$ dummy inferences are executed to allow the target and measurement setup to reach a stable operating state.

A hardware trigger is used to delimit the neural-network inference operation.
Immediately before execution of the forward pass, the firmware raises the trigger signal using \texttt{trigger\_high()}.
The trigger is lowered using \texttt{trigger\_low()} after the complete forward pass has finished.
The acquisition window starts at the rising edge of the trigger and has a fixed length.
The acquisition window therefore begins shortly before the targeted multiplication and also contains the subsequent forward-pass operations.
The analysis uses only the first $1{,}000$ samples ($250$ clock cycles) of each trace, which contain the targeted multiplication.

The host supplies the network input through a SimpleSerial \texttt{p} command.
The input value is represented as an IEEE-754 single-precision floating-point number and packed into the first four bytes of a $16$-byte command payload.
Upon receiving the command, the firmware assigns this value to the first input neuron and executes the network forward pass.

A separate SimpleSerial \texttt{k} command is used to configure the weights.
During profiling, the host sends a $32$-bit random seed to the target.
The firmware initializes its pseudo-random number generator with this seed and generates all network weights uniformly in the interval $[-2,2]$.
The generated configuration remains unchanged for a batch of $100$ consecutive measurements.
The firmware additionally returns the first weight of the first layer to the host, allowing the attacked parameter to be associated with the corresponding profiling traces.

During the attack phase, the firmware is rebuilt with the fixed target-network parameters.
The same \texttt{k} command sequence is retained in order to preserve a similar communication procedure between profiling and attack acquisition.
However, the random values generated in response to this command do not replace the fixed attack weights.
Thus, only the network input varies between attack measurements, while the unknown target weight remains constant.

The input values used in both phases are generated independently from a uniform distribution,
\begin{equation*}
    x \sim \mathcal{U}(-1,1).
\end{equation*}
A fixed host-side random seed is used to make the acquisition procedure reproducible.

The profiling and attack measurements are stored in separate ChipWhisperer projects.
The profiling project contains $50{,}000$ traces, while the attack project contains $5{,}000$ traces.
For profiling traces, the stored metadata contain both the applied input and the corresponding first-layer target weight.
For attack traces, only the applied input is stored, reflecting the assumption that the target weight is unknown during the recovery phase.

\section{Profiled Template Attack}
\label{sec:template-attack}
This section presents the proposed profiled template attack for recovering the targeted neural-network weight.
Section~\ref{sec:leakage-model} defines the floating-point leakage model, Section~\ref{sec:poi-selection} describes the selection of points of interest, and Section~\ref{sec:template-construction} explains the construction of the Gaussian templates.
Section~\ref{sec:candidate-likelihood} introduces the likelihood-based candidate scoring, while Section~\ref{sec:hierarchical-recovery} describes the coarse-to-fine-to-exact recovery procedure.

\subsection{Leakage Model}
\label{sec:leakage-model}

The proposed attack targets the intermediate value produced by the floating-point multiplication between the known input $x$ and a candidate weight $w$.
For each trace, the corresponding intermediate is modeled as
\begin{equation*}
    v = \mathrm{float32}(wx).
\end{equation*}

We assume that the instantaneous power consumption depends on the binary representation of this intermediate value.
The leakage is therefore modeled using the Hamming weight of selected bits of the IEEE-754 single-precision representation of $v$.
Let $B_m(v)$ denote the $m$ most significant bits of the IEEE-754 representation of $v$, i.e., the sign bit, the $8$ exponent bits, and the first $m-9$ fraction bits.
The leakage class at resolution $m$ is defined as
\begin{equation*}
    c_m(v)
    =
    \mathrm{HW}\left(B_m(v)\right).
\end{equation*}

To support efficient candidate refinement, three leakage resolutions are used.
The coarse model considers the sign bit, the $8$ exponent bits, and the $8$ most significant fraction bits, resulting in $17$ modeled bits and $18$ possible Hamming-weight classes.
The fine model extends the fraction contribution to $14$ bits, resulting in $23$ modeled bits and $24$ possible classes.
The exact model uses the complete $32$-bit IEEE-754 representation and therefore contains $33$ possible Hamming-weight classes.

The three leakage models can be written as
\begin{align*}
    c_{\mathrm{coarse}}(v)
    &=
    \mathrm{HW}\left(B_{17}(v)\right), \\
    c_{\mathrm{fine}}(v)
    &=
    \mathrm{HW}\left(B_{23}(v)\right), \\
    c_{\mathrm{exact}}(v)
    &=
    \mathrm{HW}\left(B_{32}(v)\right).
\end{align*}

Using fewer fraction bits in the first stages makes the model tolerant to small errors in the candidate weight.
A candidate that is close to, but not exactly equal to, the true weight produces a product whose upper bits mostly agree with the true product, while its lower fraction bits are effectively random.
The unmodeled bits still influence the measured leakage, and their contribution becomes part of the noise captured by the template covariance.
Increasing the number of modeled bits makes the model more selective, but it also requires the candidate to be closer to the true weight.
The full $32$-bit model is finally used in the exact stage, where the candidates are individual \texttt{float32} values.

During profiling, the true intermediate value is available because both $x$ and the profiling weight are known.
During the attack, the same leakage classes are computed hypothetically for each candidate weight $\hat{w}$ as
\begin{equation*}
    c_m(\hat{w},x)
    =
    \mathrm{HW}
    \left(
        B_m\left(
            \mathrm{float32}(\hat{w}x)
        \right)
    \right).
\end{equation*}
These hypothetical classes determine which profiling template is used to evaluate each candidate against the measured attack traces.

\subsection{Point-of-Interest Selection}
\label{sec:poi-selection}

The points of interest (POIs) are selected from the profiling traces by identifying time samples whose measured leakage is most strongly correlated with the targeted floating-point multiplication.
For each profiling trace $i$, the exact intermediate value
\begin{equation*}
    v_i = \mathrm{float32}(w_i x_i)
\end{equation*}
is known from the profiling input $x_i$ and profiling weight $w_i$.

A scalar leakage label is assigned to each profiling trace using the Hamming weight of the complete $32$-bit IEEE-754 representation of $v_i$,
\begin{equation*}
    h_i =
    \mathrm{HW}\left(B_{32}(v_i)\right).
\end{equation*}
For every time sample $s$ in the analysis window (the first $1{,}000$ samples of each trace), the Pearson correlation coefficient between the measured trace samples and the leakage labels is then computed as
\begin{equation*}
    \rho_s =
    \mathrm{corr}
    \left(
        \left(T_1[s],\ldots,T_N[s]\right),
        \left(h_1,\ldots,h_N\right)
    \right).
\end{equation*}

The samples are ranked according to the absolute correlation magnitude $|\rho_s|$.
To avoid selecting several samples from the same local leakage peak, a minimum temporal separation is enforced between selected POIs.
In the current implementation, five POIs are retained, with a minimum spacing of $20$ samples between them.

Let the selected positions be denoted by
\begin{equation*}
    \mathcal{P} = \{p_1,p_2,\ldots,p_5\}.
\end{equation*}
Each power trace is subsequently reduced to the five-dimensional leakage vector
\begin{equation*}
    \mathbf{t}_i =
    \left[
        T_i[p_1],
        T_i[p_2],
        T_i[p_3],
        T_i[p_4],
        T_i[p_5]
    \right].
\end{equation*}

The selected POIs are shared by the coarse, fine, and exact template sets.
Only the leakage-class definition changes between the three stages, while the same five time samples are used for template construction and candidate evaluation.

\subsection{Template Construction}
\label{sec:template-construction}

The profiling traces are grouped according to the leakage classes defined in Section~\ref{sec:leakage-model}.
Each profiling leakage vector $\mathbf{t}_i$ (Section~\ref{sec:poi-selection}) is labeled with the Hamming-weight class $c_i = c_m(v_i)$ of its known intermediate value.

For each class $c$, the class mean is estimated as
\begin{equation*}
    \boldsymbol{\mu}_c =
    \frac{1}{N_c}
    \sum_{i:c_i=c}
    \mathbf{t}_i,
\end{equation*}
where $N_c$ denotes the number of profiling traces assigned to class $c$.

Instead of estimating a separate covariance matrix for every class, the implementation uses a pooled covariance matrix shared across all classes.
This reduces the number of parameters that must be estimated and improves numerical stability for sparsely populated classes.
The pooled covariance is computed from the deviations of the profiling vectors from their corresponding class means,
\begin{equation*}
    \boldsymbol{\Sigma}
    =
    \frac{1}{N_{\mathcal{M}}}
    \sum_{c\in\mathcal{M}}
    \sum_{i:c_i=c}
    \left(
        \mathbf{t}_i-\boldsymbol{\mu}_c
    \right)
    \left(
        \mathbf{t}_i-\boldsymbol{\mu}_c
    \right)^{T},
\end{equation*}
where $\mathcal{M}$ is the set of classes with enough profiling traces for a direct estimate (see below) and $N_{\mathcal{M}}=\sum_{c\in\mathcal{M}}N_c$ is the number of profiling traces in these classes.

Each leakage class is therefore represented by a multivariate Gaussian template,
\begin{equation*}
    \mathbf{t}\mid c
    \sim
    \mathcal{N}
    \left(
        \boldsymbol{\mu}_c,
        \boldsymbol{\Sigma}
    \right).
\end{equation*}

A class mean is estimated directly only when at least $30$ profiling traces are available for that class.
Classes with fewer observations are treated as insufficiently populated.
Their mean vectors are obtained by extrapolation: for every POI, a straight line is fitted through the means of the sufficiently populated classes as a function of the Hamming weight (least squares, with each class weighted by its number of traces $N_c$), and the line is evaluated at the missing classes.
Such a class is therefore not treated as impossible; a hypothesis that requires it simply receives a low likelihood when its predicted mean is far from the measured leakage.

Three independent template sets are constructed according to the three leakage resolutions, with $18$ coarse, $24$ fine, and $33$ exact classes.
The number of classes estimated directly and by extrapolation in the recorded profiling dataset is reported in Section~\ref{sec:profiling-leakage}.

The resulting template sets are subsequently used in the corresponding stages of the hierarchical search.
The coarse templates provide robust discrimination over the complete weight interval, the fine templates refine the most likely regions, and the exact templates are used for the final \texttt{float32}-level recovery.

\subsection{Candidate Likelihood}
\label{sec:candidate-likelihood}

During the attack phase, each candidate weight is evaluated according to how well its predicted leakage classes explain the measured power traces.
For a candidate weight $\hat{w}$ and known input $x_i$, the hypothetical intermediate value is computed as
\begin{equation*}
    \hat{v}_i =
    \mathrm{float32}(\hat{w}x_i).
\end{equation*}

The corresponding leakage class is then obtained using the leakage model of the current search stage,
\begin{equation*}
    c_i(\hat{w})
    =
    \mathrm{HW}
    \left(
        B_m(\hat{v}_i)
    \right),
\end{equation*}
where $m=17$, $23$, or $32$ for the coarse, fine, or exact model, respectively.

Given the predicted class $c_i(\hat{w})$, the five-dimensional attack leakage vector $\mathbf{t}_i$ is evaluated using the corresponding Gaussian template.
The log-likelihood contribution of trace $i$ is
\begin{equation*}
    \ell_i(\hat{w})
    =
    \log
    p
    \left(
        \mathbf{t}_i
        \mid
        c_i(\hat{w})
    \right).
\end{equation*}

Using the multivariate Gaussian model, this can be expressed as
\begin{equation*}
    \ell_i(\hat{w})
    =
    -\frac{1}{2}
    \left(
        \mathbf{t}_i-\boldsymbol{\mu}_{c_i(\hat{w})}
    \right)^T
    \boldsymbol{\Sigma}^{-1}
    \left(
        \mathbf{t}_i-\boldsymbol{\mu}_{c_i(\hat{w})}
    \right)
    + K,
\end{equation*}
where $K$ contains terms that are constant with respect to the candidate weight.
Since the covariance matrix is shared across leakage classes, these constant terms do not affect candidate ranking and can be omitted during implementation.
With the constant omitted, every contribution $\ell_i(\hat{w})$ is non-positive.

Assuming that the attack traces are independent given their leakage classes, the evidence from multiple traces is accumulated by summing their log-likelihood contributions,
\begin{equation*}
    L(\hat{w})
    =
    \sum_{i=1}^{N_{\mathrm{attack}}}
    \ell_i(\hat{w}).
\end{equation*}

A correct candidate is expected to predict leakage classes that are consistently compatible with the measured traces.
Incorrect candidates produce different hypothetical multiplication results and therefore different leakage-class sequences, resulting in lower accumulated likelihoods.

The candidate with the highest accumulated score is selected as
\begin{equation*}
    \hat{w}^{\star}
    =
    \arg\max_{\hat{w}} L(\hat{w}).
\end{equation*}

The same likelihood formulation is used in all three stages of the attack.
The stages differ only in the candidate search space and in the resolution of the corresponding leakage templates.

\subsection{Hierarchical Weight Recovery}
\label{sec:hierarchical-recovery}

To avoid an exhaustive search over the complete IEEE-754 single-precision space, the proposed attack uses a hierarchical recovery procedure composed of coarse, fine, and exact stages.
Each stage evaluates candidate weights using the likelihood function introduced in Section~\ref{sec:candidate-likelihood}, while progressively increasing the resolution of both the candidate search and the leakage model.

\paragraph{Coarse search.}
The first stage searches the complete assumed weight interval
\begin{equation*}
    \hat{w} \in [-2,2]
\end{equation*}
with a step size of $10^{-4}$.
This results in $40{,}001$ candidate weights.
Each candidate is evaluated using the coarse template set based on the sign bit, exponent bits, and the $8$ most significant fraction bits of the multiplication result.
After likelihood accumulation over the available attack traces, the six highest-scoring separated peaks are retained.
A minimum separation of $0.005$ is enforced between selected peaks to avoid retaining several neighboring candidates corresponding to the same local maximum.

\paragraph{Fine search.}
Each of the six coarse finalists is refined independently within a local interval of $\pm 10^{-4}$ around the corresponding coarse estimate.
The local search uses a step size of $10^{-6}$, resulting in $201$ candidates per coarse finalist.
The candidate likelihoods are recomputed using the fine template set, which includes $14$ fraction bits in addition to the sign and exponent bits.
The best candidate from each local region is retained, producing six refined candidates.
The refined candidate with the highest fine-stage likelihood is then selected for exact recovery.

\paragraph{Exact search.}
The best fine candidate is interpreted directly as an IEEE-754 single-precision value.
Its neighboring representable \texttt{float32} values are enumerated within $\pm512$ units in the last place (ULPs).
These values are additionally evaluated with both signs and scaled by $2^j$ for $j\in\{-10,\ldots,10\}$, and candidates outside $[-2,2]$ are discarded.
The scaled variants $\pm\hat{w}\,2^j$ have exactly the same fraction bits as $\hat{w}$, so their products differ from the true product only in the sign and exponent bits.
The coarse and fine models cannot separate them reliably (Section~\ref{sec:weight-recovery}), and the exact stage must be able to reach the true weight even if the fine stage selected one of these look-alikes.
For a fine result close to $1.43$, this gives $1{,}025$ neighboring values, $11$ scales that stay within $[-2,2]$, and $2$ signs, i.e., $22{,}550$ exact candidates.

Each remaining candidate is evaluated using the exact template set based on the complete $32$-bit representation of the hypothetical multiplication result,
\begin{equation*}
    c_i(\hat{w})
    =
    \mathrm{HW}
    \left(
        B_{32}\left(\mathrm{float32}(\hat{w}x_i)\right)
    \right).
\end{equation*}
The candidate with the highest accumulated exact-template likelihood is returned as the recovered weight,
\begin{equation*}
    \hat{w}^{\star}
    =
    \arg\max_{\hat{w}\in\mathcal{W}_{\mathrm{exact}}}
    L_{\mathrm{exact}}(\hat{w}).
\end{equation*}

The hierarchical procedure therefore reduces the search space progressively.
The coarse stage identifies promising regions over the full weight interval, the fine stage resolves nearby numerical candidates, and the exact stage distinguishes neighboring IEEE-754 representations and power-of-two look-alikes to obtain a bit-precise \texttt{float32} estimate.

\section{Experimental Results}
\label{sec:evaluation}

\subsection{Profiling Leakage}
\label{sec:profiling-leakage}

The profiling traces exhibit clear data-dependent leakage associated with the targeted floating-point multiplication.
Using the full $32$-bit Hamming-weight model defined in Section~\ref{sec:leakage-model}, the correlation analysis described in Section~\ref{sec:poi-selection} identifies several distinct leakage peaks within the analysis window.

The five selected points of interest are located at sample indices
\begin{equation*}
    \mathcal{P} =
    \{518,\,569,\,622,\,657,\,737\}.
\end{equation*}
The correlation is negative at all five POIs, and its absolute value lies between $0.74$ and $0.93$.
This indicates a strong statistical dependence between the measured power consumption and the Hamming weight of the floating-point multiplication result.

\begin{figure}[htb]
    \centering
    \includegraphics[width=0.7\linewidth]{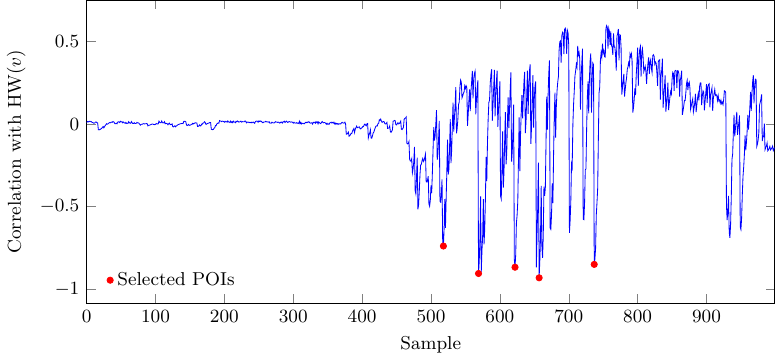}
    \caption{Pearson correlation between the profiling traces and the Hamming weight of the full $32$-bit product $v=\mathrm{float32}(wx)$ for every sample of the analysis window. The red markers are the five selected POIs.}
    \label{fig:profiling-correlation}
\end{figure}

Figure~\ref{fig:profiling-correlation} shows the correlation for every sample of the analysis window together with the selected POIs.
The first $\approx 400$ samples show almost no correlation, which is consistent with the operations executed between the trigger and the start of the multiplication.
The several strong and temporally separated peaks after that suggest that the intermediate value influences the measured leakage at multiple points during the software multiplication routine and the operations that follow it, such as storing the product and adding it to the neuron's sum.

The profiling data are subsequently divided into Hamming-weight classes for construction of the coarse, fine, and exact templates.
Not all theoretically possible classes are sufficiently represented in the profiling dataset.
By using a minimum threshold of $30$ traces per class, $11$ of the $18$ coarse classes, $13$ of the $24$ fine classes, and $16$ of the $33$ exact classes are estimated directly from measured traces.
The remaining class means are obtained through the extrapolation procedure described in Section~\ref{sec:template-construction}.

\begin{figure}[htb]
    \centering
    \includegraphics[width=0.32\linewidth]{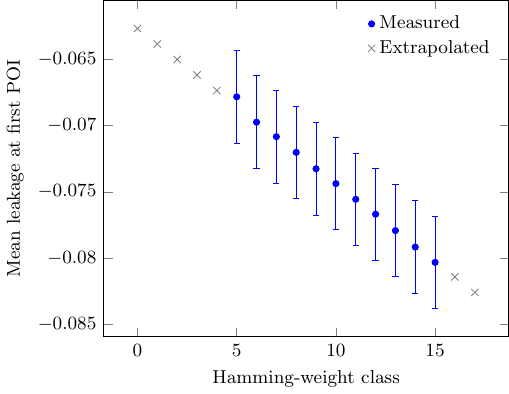}\hfill
    \includegraphics[width=0.32\linewidth]{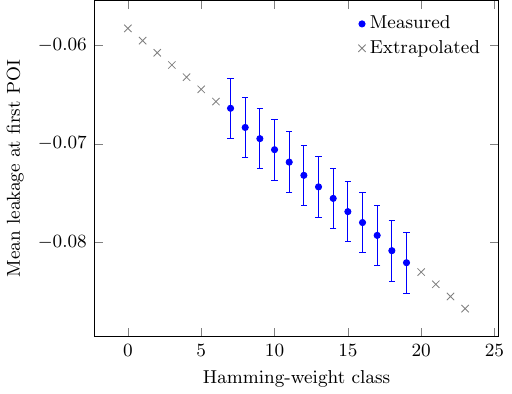}\hfill
    \includegraphics[width=0.32\linewidth]{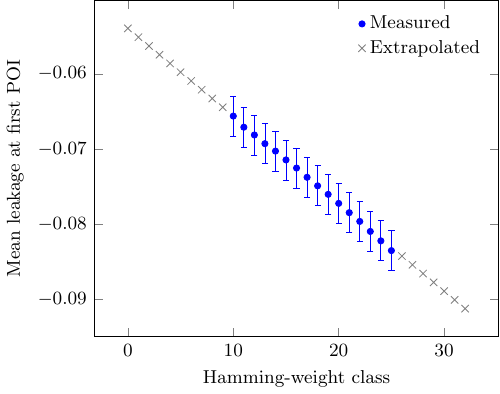}
    \caption{Template means at the first POI for the coarse (left), fine (middle), and exact (right) template sets. Blue: classes estimated from at least $30$ profiling traces, with error bars of one standard deviation of the noise. Gray: extrapolated classes.}
    \label{fig:templates}
\end{figure}

Figure~\ref{fig:templates} shows the resulting class means at the first POI.
For all three template sets, the means of the measured classes decrease almost linearly with the Hamming weight, which supports the linear extrapolation of the missing classes.
The error bars show the standard deviation of the noise at this POI.
It is two to three times larger than the difference between neighboring class means, so a single trace carries little information and evidence from many traces has to be combined.

These results confirm that the targeted multiplication produces sufficiently strong leakage for template construction, while also illustrating the increasing sparsity of Hamming-weight classes as more IEEE-754 bits are included in the leakage model.

\subsection{Weight Recovery}
\label{sec:weight-recovery}

We evaluate the proposed attack on the fixed target network described in Section~\ref{sec:target-network}.
The configured value of the attacked weight is $1.43$, which is represented during execution by its nearest IEEE-754 single-precision value,
\begin{equation*}
    w^\star \approx 1.4299999.
\end{equation*}

The hierarchical recovery procedure described in Section~\ref{sec:hierarchical-recovery} is applied to the first $1{,}000$ attack traces.
The six coarse finalists are $1.4299$, $0.7150$, $0.3575$, $0.0894$, $-1.4300$, and $0.1787$.
Apart from the true weight, these are its look-alikes $w^\star/2$, $w^\star/4$, $w^\star/16$, $-w^\star$, and $w^\star/8$, whose products share the fraction bits of the true product.
The fine stage ranks the finalist near the true weight highest and refines it to $1.4299961$, which is $33$~ULPs below the true value.
The exact stage then recovers
\begin{equation*}
    \hat{w} = 1.4299999,
\end{equation*}
which matches the true single-precision representation of the target weight bit for bit (\texttt{0x3FB70A3D}).
The same exact value is recovered when all $5{,}000$ available attack traces are used.

The recovery error is therefore
\begin{equation*}
    |\hat{w} - w^\star| = 0,
\end{equation*}
when evaluated in the \texttt{float32} representation used by the implementation.

These results demonstrate that the proposed profiled template attack is able to identify the exact single-precision representation of the targeted first-layer weight from the measured power traces.
The following sections further analyze how candidate likelihoods evolve as traces are accumulated and how the probability of successful recovery depends on the number of attack traces.

\subsection{Candidate Convergence}
\label{sec:candidate-convergence}

To illustrate how evidence accumulates across attack traces, we analyze the cumulative log-likelihood of the candidates of every stage.
For a candidate weight $\hat{w}$ and the first $n$ attack traces, the cumulative score is defined as
\begin{equation*}
    L_n(\hat{w})
    =
    \sum_{i=1}^{n}
    \ell_i(\hat{w}),
\end{equation*}
where $\ell_i(\hat{w})$ is computed with the templates of the respective stage.

\begin{figure}[!htbp]
    \centering
    \begin{tabular}{cc}
        \includegraphics[width=0.5\linewidth]{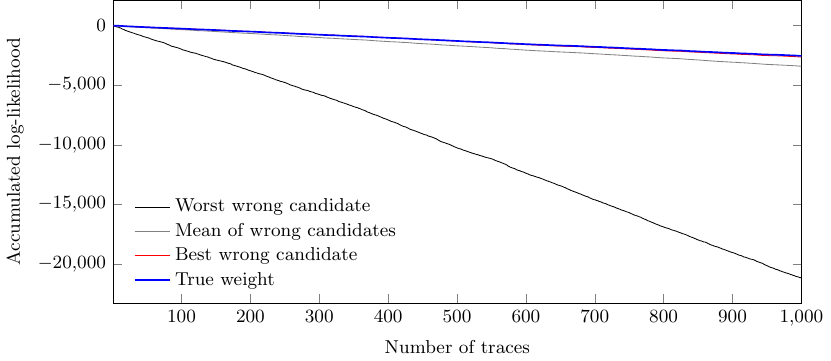} &  \includegraphics[width=0.5\linewidth]{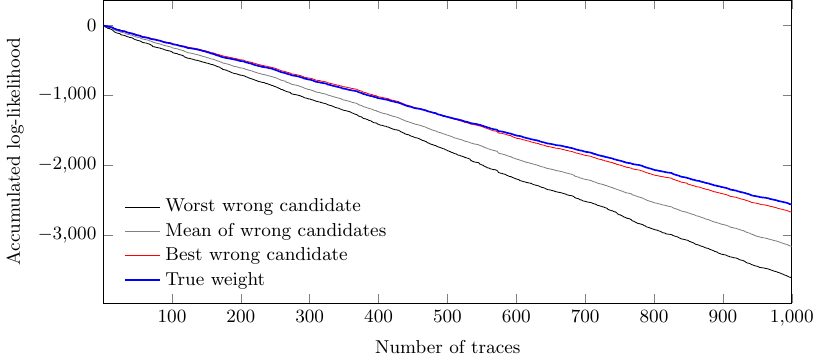}\\
        \multicolumn{2}{c}{\includegraphics[width=0.5\linewidth]{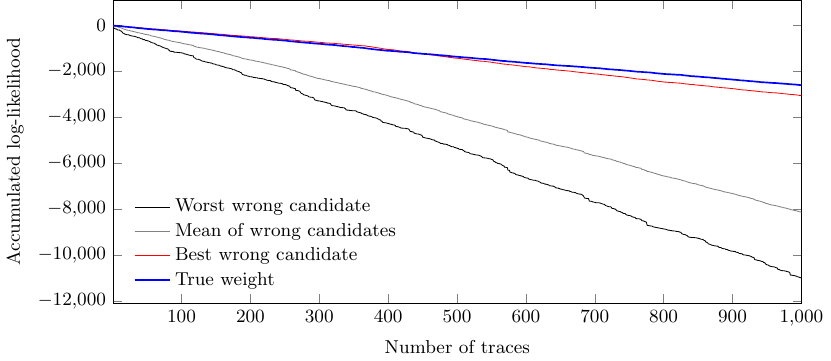}} 
    \end{tabular}
    \caption{Cumulative log-likelihood of the true weight and of the best, mean, and worst wrong candidate over the first $1{,}000$ attack traces, for the coarse (top left, $39{,}900$ wrong candidates), fine (top right, $1{,}005$), and exact (bottom, $22{,}549$) stage.}
    \label{fig:candidate-convergence}
\end{figure}

Figure~\ref{fig:candidate-convergence} shows $L_n$ for the first $1{,}000$ attack traces in acquisition order.
For every stage, the true weight is compared with all wrong candidates of that stage: the complete coarse grid, the fine grids around all six coarse finalists, and all exact candidates.
Instead of drawing every wrong candidate, the figure shows the best, the mean, and the worst score among them for every $n$.
In the coarse and fine stages, candidates within $0.005$ of the true weight are counted as correct, since these stages only have to reach the neighborhood of the true weight.
In the exact stage, every candidate except the bit-exact true weight is counted as wrong.

Since every contribution $\ell_i(\hat{w})$ is non-positive (Section~\ref{sec:candidate-likelihood}), all curves decrease as more traces are added.
Candidate ranking is determined by the relative score, with the most plausible candidate corresponding to the highest, i.e., least negative, cumulative log-likelihood.
The mean and the worst wrong candidate fall behind the true weight quickly: after $10$ traces, about $97\%$ or more of the wrong candidates of every stage already score below the true weight.
The best wrong candidate stays close to the true weight much longer: the true weight scores higher than every wrong candidate after $427$ traces in the coarse stage, $497$ traces in the fine stage, and $446$ traces in the exact stage.
In the exact stage, all $1{,}024$ neighboring \texttt{float32} values within $\pm512$~ULPs are already outscored after $7$ traces.
The strongest competitors are the power-of-two look-alikes ($w^\star/2$, $w^\star/4$, $w^\star/16$, and $w^\star/256$), whose products differ from the true product only in the exponent bits.

The coarse and fine stages do not need the true weight to score highest overall.
The coarse stage only needs it among the six finalists, and the fine stage only compares the six refined finalists.
The trace order shown here is also harder than average: the complete attack on these traces selects the look-alike $w^\star/256$ until $446$ traces, whereas random subsets of $73$ traces already succeed in $88\%$ of the cases (Section~\ref{sec:success-rate}).

This convergence behavior also motivates the success-rate analysis in Section~\ref{sec:success-rate}, where the attack is repeated for different numbers of traces to quantify the probability of recovering the correct weight.

\subsection{Success Rate}
\label{sec:success-rate}

To quantify the number of attack traces required for reliable recovery, we evaluate the empirical success rate of the complete hierarchical attack for increasing trace counts.
The trace counts $N$ are $26$ values spaced logarithmically between $1$ and $500$.
The experiment is repeated $100$ times.
In every repetition, $500$ of the $5{,}000$ attack traces are drawn in a random order, and the attack is run on the first $N$ of them for every trace count, so the subsets are nested.
A recovery is considered successful only when the recovered
 value equals the true \texttt{float32} weight bit for bit,
 i.e., $\hat{w}=w^\star$, the same criterion as in Section~
\ref{sec:weight-recovery}.

The empirical success rate is defined as
\begin{equation*}
    \mathrm{SR}(N)
    =
    \frac{
        \#\{\text{successful recoveries using }N\text{ traces}\}
    }{100}.
\end{equation*}

\begin{figure}[htb]
    \centering
    \includegraphics[width=0.6\linewidth]{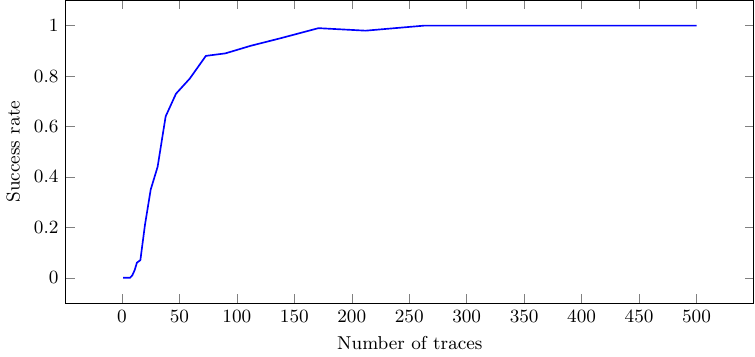}
    \caption{Empirical success rate of the proposed weight-recovery attack as a function of the number of attack traces.}
    \label{fig:success-rate}
\end{figure}

Figure~\ref{fig:success-rate} shows the resulting success rate as a function of the number of attack traces.
The attack exhibits a rapid improvement as additional traces are included, reaching approximately $64\%$ success with $38$ traces, $88\%$ with $73$ traces, and $95\%$ with $138$ traces.
The success rate reaches $99\%$ with $171$ traces and $100\%$ from $263$ traces up to the largest evaluated count of $500$.

These results indicate that the proposed profiling attack does not require the full set of $5{,}000$ attack traces to recover the target weight reliably.
Instead, the likelihood information accumulated from a few hundred measurements is sufficient to obtain consistent recovery in the evaluated setting.

\section{Discussion}
\label{sec:discussion}
\textbf{Interpretation of the Results} The results indicate that the profiling model captures leakage associated with the targeted floating-point multiplication rather than with a single fixed network configuration.
Although the non-target weights vary during profiling and are not included in the leakage model, the resulting templates remain sufficiently discriminative to recover the target weight in the fixed attack network.
The success-rate analysis further shows that evidence accumulates consistently in favor of the correct candidate as more traces are used.
This supports the assumption that the selected POIs and templates provide a meaningful statistical model of the leakage of
\begin{equation*}
    \mathrm{float32}(wx).
\end{equation*}

\noindent
\textbf{Role of the Hierarchical Search}
The hierarchical search balances robustness and precision during weight recovery.
The coarse stage uses a reduced IEEE-754 leakage model to identify promising regions over the full weight interval, while the fine stage increases the leakage resolution to distinguish nearby candidates.
The final stage then evaluates neighboring \texttt{float32} values using the complete $32$-bit leakage model.
This progression avoids applying the most detailed leakage model to the entire candidate space.
It therefore reduces the computational cost of the search while preserving the ability to recover the exact floating-point representation of the target weight.

\noindent
\textbf{Limitations and Future Work}
The present study focuses on the first weight of the first layer on a single target platform, providing an initial evaluation of the proposed profiling methodology.
A natural next step is to extend the analysis to multiple network configurations and deeper layers.
Future work should also investigate the portability of the templates across different devices and measurement sessions, as well as the effect of larger or more balanced profiling datasets on sparsely populated Hamming-weight classes.

\section{Conclusion}
\label{sec:conclusion}

This work presented a profiled template attack for recovering a single-precision neural-network weight from power side-channel measurements.
The attack models the leakage of the floating-point multiplication between a known input and the unknown target weight, and uses templates constructed from randomized profiling configurations.
A hierarchical coarse-to-fine-to-exact search enables efficient exploration of the candidate space and exact recovery of the target \texttt{float32} value.
The experimental evaluation demonstrates successful weight recovery and shows that reliable recovery can be achieved with only a few hundred attack traces in the considered setting.

\bibliographystyle{splncs04}
\bibliography{bib}

\end{document}